\documentclass[aps,superscriptaddress,twocolumn,twoside,floatfix,nofootinbib]{revtex4-1}
\usepackage{graphicx,amsmath,amssymb,color,array}
\usepackage[normalem]{ulem}
\usepackage{multirow}
\usepackage{times}
\usepackage[ocgcolorlinks,colorlinks=true,linkcolor=blue,citecolor=red,linktocpage=true]{hyperref}

\usepackage{amsfonts}
\usepackage{latexsym}
\usepackage{amsmath}
\usepackage{amssymb}
\usepackage{amsfonts}
\usepackage{amsthm}
\usepackage{mathrsfs}
\usepackage{natbib}
\usepackage{color,verbatim,graphics}
\usepackage{psfrag}
\DeclareMathAlphabet{\mathrsfs}{U}{rsfs}{m}{n}
\DeclareMathAlphabet{\mathpzc}{OT1}{pzc}{m}{it}
\DeclareMathAlphabet{\matheus}{U}{eus}{m}{n}
\DeclareMathAlphabet{\mathbbold}{U}{bbold}{m}{n}
\DeclareMathOperator*{\argmin}{arg\,min}

\newcommand{\ba}{\begin{eqnarray}}
\newcommand{\be}{\begin{equation}}
\newcommand{\ee}{\end{equation}}
\newcommand{\beq}{\begin{equation}}
\newcommand{\eeq}{  \end{equation}}
\newcommand{\bea}{\begin{eqnarray}}
\newcommand{\eea}{  \end{eqnarray}}
\newcommand{\nc}[0]{\mathrm{\o}}

\newcommand{\ea}{\end{eqnarray}}
\newcommand{\ban}{\begin{eqnarray*}}
\newcommand{\ean}{\end{eqnarray*}}
\newcommand{\Tr}{\operatorname{tr}}

\newcommand{\ket}[1]{|#1\rangle}
\newcommand{\bra}[1]{\langle#1|}

\newcommand{\ketbra}[2]{|#1\rangle\langle#2|}

\newcommand{\eg}{{\it{e.g.~}}}
\newcommand{\ie}{{\it{i.e.}~}}
\newcommand{\etal}{{\it{et al. }}}

\newcommand{\rA}{\mathrm{A}}
\newcommand{\rB}{\mathrm{B}}

\begin{document}

\title{Quantatitive relations between measurement incompatibility, quantum steering, and nonlocality}

\author{D. Cavalcanti}
\affiliation{ICFO-Institut de Ciencies Fotoniques, The Barcelona Institute of Science and Technology, 08860 Castelldefels (Barcelona), Spain}
\email{daniel.cavalcanti@icfo.es}
\author{P. Skrzypczyk}
\affiliation{H. H. Wills Physics Laboratory, University of Bristol, Tyndall Avenue, Bristol, BS8 1TL, United Kingdom}
\begin{abstract}
The certification of Bell nonlocality or quantum steering implies the use of incompatible measurements. Here we make this connection quantitative. We show how to strengthen robustness-based steering and nonlocality quantifiers in order that they give strong lower bounds to previously proposed incompatibility quantifiers. Our results can be seen from two perspectives. On the one hand, they can be used to estimate how much steering or nonlocality can be demonstrated with a given set of measurements. On the other hand, they gives one-sided device-independent and device-independent ways of estimating measurement incompatibility.
\end{abstract}

 \maketitle

\section{Introduction}

Some measurements in quantum mechanics cannot be performed simultaneously. The most standard notion of \textit{measurement incompatibility} is non commutativity, which lies at the heart of the so-called Robertson-Heisenberg uncertainty principle \cite{UncPrin}. Although commutativity characterises well the incompatibility of projective measurements, it does not account properly for the incompatibility of general positive-operator-valued-measure (POVM) measurements. In this case, a better suited figure of merit is joint measurability. A set of measurements $\{M_x\}$ is said to be jointly measurable if there exists a parent POVM measurement $N$, from which each measurement $M_x$ can be derived from $N$ by coarse graining \cite{Busch}.

Characterising measurement incompatibility is an important task, since it gives a notion of nonclassicality for quantum measurements and is behind quantum information applications such as quantum key distribution \cite{Nielsen}. Measurement incompatibility is also closely related to the notion of non-classical correlations. It is well known that jointly measurable measurements cannot lead to the violation of Bell inequalities \cite{Bell,review} or quantum steering \cite{Sch,Wiseman}. Thus, any Bell inequality violation or quantum steering observation certifies the use of incompatible measurements. The converse of this result is however less trivial. Whether any set of non jointly measurable measurements can lead to the violation of a Bell inequality is still an open question, despite recent progress \cite{JMvsNLWolf,JMvsNLGeneva}. Notice however that a strict connection between joint measurability and quantum steering has indeed been proven: Any set of non jointly measurable measurements can be used to demonstrate quantum steering \cite{JMSteering1,JMSteering2}.

While the quantification of nonlocal correlations has been well studied (see \eg \cite{review,vicente} and references therein), only more recently has the quantification of both quantum steering and measurement incompatibility been addressed \cite{QuantSteeringUs,QuantSteeringPiani,QuantJM,PuseyJOSA,Uola}. In these works quantum steering and joint measurability were quantified through their robustness to noise (see also Refs.~\cite{ResourceSteering,HeinosaariReview} for a more general approach to steering and measurement incompatibility  quantification). In the present paper we show that, in any steering or Bell test, the associated quantifiers satisfy the following relation:
\be \label{e:relation}
\mathrm{I} \geq \mathrm{S} \geq \mathrm{NL},
\ee
where $\mathrm{I}$, $\mathrm{S}$, and $\mathrm{NL}$ refer to the quantifiers of measurement incompatibility, steering and nonlocality respectively. This shows that, for a known set of measurements, upper bounds on the amount of steering and nonlocality that can be demonstrated can be obtained. Furthermore, we propose new robustness-based nonlocality and quantum steering quantifiers that are strengthened versions of their standard counterparts, such that relation \eqref{e:relation} still holds. This strengthened relation allows us to estimate how incompatible a set of measurements is in a one-sided device-independent way (with the knowledge of $\mathrm{S}$) or in a fully device-independent way (with the knowledge of $\mathrm{NL}$). We demonstrate the merit of these strengthened quantifiers by proving that in the case that Alice and Bob share a full Schmidt-rank pure entangled state, then the one-sided estimation is precise, \ie Bob estimates the exact amount of incompatibility in Alice's measurements in the steering scenario.

This paper is organised as follows. We first review the definition of joint measurability and the proposals for quantifying it (Sec.~\ref{sec: JM}). We then discuss the concepts of quantum steering and Bell nonlocality and their quantifications (Secs.~\ref{sec: steering} and \ref{sec: nonlocality}). In Sec.~\ref{sec: estimating} we show how the amount of steering and nonlocality a set of measurements can demonstrate is bounded by measurement incompatibility. In Secs.~\ref{sec: 1sided} and \ref{sec: DI} we present modified quantifiers of quantum steering and Bell nonlocality and show that they provide tighter one-sided device-independent and device-independent lower bounds on the amount of measurement incompatibility, respectively. In Sec.~\ref{sec: pure states} we show the tightness of the previous results in the case of quantum steering when pure entangled states are shared between the parties. We finally illustrate our findings with theoretical examples and also analyse recent loophole-free experimental tests of steering and nonlocality in Sec.~\ref{sec: examples}. We summarise in Sec.~\ref{sec: conclusion}.

\section{Measurement incompatibility and its quantification}\label{sec: JM}

Two measurements $\mathbb{M}_0=\{M_{a|0}\}_a$ and $\mathbb{M}_1=\{M_{a'|1}\}_{a'}$, with measurement operators $M_{a|0}$ and $M_{a'|1}$ respectively, are said to be compatible (or jointly measurable) if there exists a parent measurement $\mathbb{G}=\{G_{aa'}\}_{aa'}$ that works as a refinement for both of them, in the sense that they can be obtained from $\mathbb{G}$ via coarse graining. Formally, this means that the measurement operators of $\mathbb{M}_0$ and $\mathbb{M}_1$ are obtained from those of $\mathbb{G}$ by
\begin{equation}
M_{a|0}=\sum_{a'}G_{aa'}~\text{and}~M_{a'|1}=\sum_{a}G_{aa'}.
\end{equation}
If no such parent measurement can be found it is because $\mathbb{M}_0$ and $\mathbb{M}_1$ cannot be measured simultaneously, and they are then called incompatible.

The idea of incompatibility can be extended to an arbitrary number of measurements. Suppose a set of $m$ measurements $\mathcal{M}=\{\mathbb{M}_x\}_{x=1}^m$, each one composed by measurement operators $\{M_{a|x}\}_{a=1}^n$, corresponding to each of the $n$ outcomes\footnote{Without loss of generality we assume that all measurement $x$ have the same number of outcomes. If two measurements do not have the same number of outcomes we can extend the one with fewer outcomes by adding null operators, which represent the outcomes that never occur.}, such that $\sum_{a=1}^n M_{a|x}=\openone$ $\forall x$, and $M_{a|x}\geq0$ $\forall a,x$. The set $\mathcal{M}$ is said to be jointly measurable if there exists a parent measurement $\mathbb{G}$ defined by measurement operators $G_{\vec{a}}$ whose outcomes are labeled by $\vec{a} = (a_{x=1}, a_{x=2}, . . . , a_{x=m})$ such that
\begin{equation}
G_{\vec{a}}\geq 0,\quad
\sum_{\vec{a}}
G_{\vec{a}} = \openone, \quad
\sum_{\vec{a}\setminus a_x}
G_{\vec{a}}= M_{a_x|x},
\end{equation}
where the $a_x$ refer to the measurement outcome of the measurement $x$ and $\vec{a}\setminus a_x$ stands for the elements of $\vec{a}$ except for ${a_x}$.
This definition says that all measurements in the set $\mathcal{M}$ can be measured simultaneously by performing the measurement $G$ and appropriately grouping outcomes.

In Refs.~\cite{QuantJM,Uola,PuseyJOSA} several robustness-based quantifiers of joint measurability were defined. The idea is to quantify incompatibility by how much noise has to be added to the measurements such that they become jointly measurable. Depending on the type of noise considered, different quantifiers are defined as we recall below.

In what follows we will abuse notation and refer to a set of measurements simply as $M_{a|x}$.
\subsection{Incompatibility robustness}
One of these quantifiers is the \textit{Incompatibility Robustness} of a set of measurements $M_{a|x}$, proposed in Ref.~\cite{Uola}. It is defined as the minimal $t$ such that there exist another set of measurements $N_{a|x}$ for which the mixture $(\mathbb{M}_{x}+t\mathbb{N}_{x})/(1+t)$ is jointly measurable. Mathematically
\begin{align}\label{IR}
\mathrm{IR}(M_{a|x}) = \min\,& t\\
\text{s.t. }  &t\geq0,\nonumber\\
&\frac{M_{a_x|x}+tN_{a_x|x}}{1+t}=\sum_{\vec{a}\setminus a_x}
G_{\vec{a}} \:\: \forall x,a_x,\nonumber\\
&N_{a|x}\geq0\:\:\forall a,x,\quad\sum_{a}N_{a|x}=\openone\:\:\forall x,\nonumber\\
&G_{\vec{a}}\geq 0\:\:\forall\vec{a},\quad\sum_{\vec{a}}
G_{\vec{a}} = \openone\nonumber
\end{align}

\subsection{Incompatibility Random Robustness}

Another previously defined quantifier of measurement incompatibility is the \textit{Incompatibility Random Robustness} \cite{QuantJM}, which is in fact a special case of the Incompatibility Robustness, when $N_{a|x}=\openone/n$, (where $n$ is the number of outcomes the measurements $M_{a|x}$ have). In this case we have
\begin{align}\label{rIR}
\mathrm{IR}^\mathrm{r}(M_{a|x})=\min\, &t\\
\text{s.t. } & t\geq0,\nonumber\\
&\frac{M_{a_x|x}+t\openone/n}{1+t}=\sum_{\vec{a}\setminus a_x}G_{\vec{a}} \:\: \forall x,a_x,\nonumber\\
&G_{\vec{a}}\geq 0\:\:\forall\vec{a},\quad\sum_{\vec{a}}
G_{\vec{a}} = \openone\nonumber.
\end{align}

\subsection{Incompatibility Jointly-Measurable Robustness}

Similarly to above, we can also define the \textit{Jointly-Measurable Robustness}, for which the noise $N_{a|x}$ is given by jointly-measurable measurements. In this case we have
\begin{align}\label{jmIR}
\mathrm{IR}^\mathrm{jm}(M_{a|x})=\min\, &t\\
\text{s.t. } & t\geq0,\nonumber\\
&\frac{M_{a_x|x}+tN_{a_x|x}}{1+t}=\sum_{\vec{a}\setminus a_x}G_{\vec{a}} \:\: \forall x,a_x,\nonumber\\
&G_{\vec{a}}\geq 0\:\:\forall\vec{a},\quad \sum_{\vec{a}}G_{\vec{a}} = \openone\nonumber\nonumber\\
&N_{a_x|x}=\sum_{\vec{a}\setminus a_x}H_{\vec{a}} \:\: \forall x,a_x,\nonumber\\
&H_{\vec{a}}\geq 0\:\:\forall\vec{a},\quad\sum_{\vec{a}}
H_{\vec{a}} = \openone\nonumber
\end{align}

\subsection{Incompatibility Weight}

Finally, the \textit{Incompatibility Weight} of a set of measurements $M_{a|x}$ was defined in \cite{PuseyJOSA}. It is based on a decomposition of the measurements $M_{a|x}$ in terms of a convex combination of an arbitrary set of measurement $O_{a|x}$ and an arbitrary set of jointly-measurable measurements $N_{a|x}$. The Incompatibility Weight is the maximal weight of the jointly measurable set $N_{a|x}$ that can be used in such a decomposition.
\begin{align}\label{IW}
\mathrm{IW}(M_{a|x})=\min\, &t\\
\text{s.t. }&t\geq0,\nonumber\\
& M_{a|x}=tO_{a|x}+(1-t)N_{a|x}\:\: \forall x,a,\nonumber\\
& O_{a|x}\geq0\:\:\forall a,x,\quad \sum_{a} O_{a|x}=\openone\:\:\forall x,\nonumber\\
& N_{a_x|x}=\sum_{\vec{a}\setminus a_x}G_{\vec{a}}\:\:\forall x,a_x,\nonumber\\
& G_{\vec{a}}\geq 0\:\:\forall\vec{a},\quad\sum_{\vec{a}}
G_{\vec{a}} = \openone\nonumber.\nonumber
\end{align}

It has been shown that each of the four quantifiers can be re-expressed in the form of a semidefinite program (SDP), a class of convex optimization problems that can be solved efficiently \cite{sdp}. Thus, each of these quantifiers can easily be calculated in many simple cases of interest, using standard software packages.

\section{Quantum steering and its quantification}\label{sec: steering}

Quantum steering refers to a bipartite situation where two parties, Alice and Bob, share an unknown bipartite state $\rho_{\mathrm{A}\mathrm{B}}$ onto which Alice performs unknown measurements $M_{a|x}$, labelled by $x$. Bob, in turn, can perform tomography on his system and determine what are the conditional states he is left with, after Alice performs a given measurement $x$ and obtains an outcome $a$. In mathematical terms, Bob will observe a collection of post-measured states defined by
\begin{equation}\label{assemblage}
\sigma_{a|x}=\Tr_\mathrm{A}[(M_{a|x}\otimes\openone) \rho_{\mathrm{A}\mathrm{B}}],
\end{equation}
where the measurement operators of $\mathbb{M}_x$ are $\{M_{a|x}\}_a$. Notice that the states \eqref{assemblage} are not normalised, with their normalisation defined by $\Tr [\sigma_{a|x}]=P(a|x)$, \ie the probability that Alice obtains outcome $a$ after she chooses measurement $x$. The collection of ensembles $\{\sigma_{a|x}\}_{a,x}$ is often called an \emph{assemblage} \cite{Pusey}. Note that we will exclusively consider the situation here where Alice steers Bob (as is customary), however all results could be equally re-derived in the converse direction, where Bob steers Alice. In what follows, we will abuse notation denote an assemblage simply as $\sigma_{a|x}$. 

An assemblage is said to demonstrate steering if it does not admit a decomposition of the form \cite{Pusey}
\begin{equation}\label{LHS}
\sigma_{a|x}=\sum_\lambda D(a|x,\lambda)\sigma_\lambda \quad \forall x,a,
\end{equation}
where $D(a|x,\lambda)$ are deterministic probability distributions assigning one particular outcome for each measurement $x$, and the operators $\sigma_\lambda$ satisfy $\sigma_\lambda\geq0$ and $\sum_\lambda \Tr [\sigma_\lambda]=1$. The decomposition \eqref{LHS} is called a local-hidden-state (LHS) model \cite{Wiseman}, and any assemblage satisfying \eqref{LHS} is said to be a LHS assemblage.

For a finite number of inputs $x$ and outcomes $a$ there exists a finite number of deterministic probability distributions $\{D(a|x,\lambda)\}_a$. Combined with the fact that \eqref{LHS} is a linear matrix inequality, it can be seen that deciding if a given assemblage demonstrates steering can also be solved via semi-definite programming \cite{Pusey,SteeringReview}.

\subsection{Steering Robustness}
The \textit{Steering Robustness} ($\mathrm{SR}$) \cite{QuantSteeringPiani} quantifies the minimal amount of (arbitrary) noise $\pi_{a|x}$ that one has to add to an assemblage $\sigma_{a|x}$ such that their mixture is a LHS assemblage. More precisely
\begin{align}\label{SR}
\mathrm{SR}(\sigma_{a|x})=\min\, &s\\
\text{s.t. } &\frac{\sigma_{a|x}+s\pi_{a|x}}{1+s}=\sum_\lambda D(a|x,\lambda)\sigma_\lambda\:\: \forall a,x,\nonumber\\
&s\geq0,\quad \pi_{a|x}\geq0\:\:\forall a,x,\nonumber \\
&\sigma_\lambda\geq0\:\: \forall \lambda, \quad \Tr\sum_\lambda \sigma_\lambda = 1.
\end{align}

\subsection{Reduced-state Steering Robustness}
Along the same line of reasoning as above, we now define a steering quantifier which we call the \emph{Reduced-state Steering Robustness}. The starting point is the Steering Robustness, but now instead of adding arbitrary noise, we restrict to `reduced-state' noise -- whereby $\pi_{a|x} = \rho_\mathrm{B}/n$ $\forall a,x$. More precisely,
\begin{align}\label{SRred}
\mathrm{SR}^\mathrm{red}(\sigma_{a|x})=\min\, &s\\
\text{s.t. } &\frac{\sigma_{a|x}+s\rho_\mathrm{B}/n}{1+s}=\sum_\lambda D(a|x,\lambda)\sigma_\lambda\:\: \forall a,x,\nonumber \\
&\rho_\mathrm{B} = \sum_{a}\sigma_{a|x} \:\:\forall x,\nonumber\\
&s\geq0,\quad \sigma_\lambda\geq0\:\: \forall \lambda\nonumber
\end{align}
This quantity quantifies the minimum amount of reduced state that can be added to an assemblage such that the final assemblage becomes LHS.

\subsection{LHS Steering Robustness}
Similarly as before the \emph{LHS Steering Robustness} \cite{SABGS} is a modification of the Steering Robustness, where the noise added is now given by LHS noise (\ie by adding a LHS assemblage). More precisely,
\begin{align}\label{LHSR}
\mathrm{SR}^\mathrm{lhs}(\sigma_{a|x})=\min\, &s\\
\text{s.t. } &\frac{\sigma_{a|x}+s\gamma_{a|x}}{1+s}=\sum_\lambda D(a|x,\lambda)\sigma_\lambda\:\: \forall a,x,\nonumber \\
&\gamma_{a|x}=\sum_\lambda D(a|x,\lambda)\gamma_\lambda\:\: \forall a,x,\nonumber \\
&s\geq0,\quad \gamma_\lambda\geq0\:\: \forall \lambda,\nonumber\\
& \sigma_\lambda \geq 0\:\: \forall \lambda, \quad \Tr\sum_\lambda \sigma_\lambda = 1.
\end{align}
This quantity gives the minimum amount of LHS assemblage that has to be added to a generic assemblage such that the resulting one becomes LHS.

\subsection{Steering Weight}

The last steering quantifier we work with is the \textit{Steering Weight} ($\mathrm{SW}$) \cite{QuantSteeringUs}. It consists in decomposing an assemblage into a convex combination of a generic assemblage $\pi_{a|x}$ and another assemblage with a LHS model $\gamma_{a|x}$, and then asking how much weight we can put on the component $\gamma_{a|x}$. Formally it is defined by:
\begin{align}\label{SW}
\mathrm{SW}(\sigma_{a|x})=\min\, &s\\
\text{s.t. } &\sigma_{a|x}=s\pi_{a|x}+(1-s)\gamma_{a|x},\nonumber\\
& \gamma_{a|x}=\sum_\lambda D(a|x,\lambda)\sigma_\lambda\:\:\forall a,x\nonumber\\
&s\geq0,\quad \pi_{a|x}\geq0\:\:\forall a,x\nonumber \\
& \sigma_\lambda\geq0 \:\:\forall \lambda,\quad \Tr\sum_\lambda\sigma_\lambda = 1.\nonumber
\end{align}

\section{Bell nonlocality and its quantification}\label{sec: nonlocality}

Bell nonlocality (in a bipartite scenario) refers to the situation where Bob also performs unknown measurements $M'_{b|y}$, labelled by $y$, on the shared unknown state $\rho_\mathrm{AB}$. In mathematical terms, Alice and Bob will observe a collection of probability distributions defined by
\begin{equation}\label{lhv}
P(ab|xy) = \Tr[(M_{a|x}\otimes M'_{b|y})\rho_\mathrm{AB}].
\end{equation}
The collection of probability distributions $\{P(ab|xy)\}_{a,b,x,y}$ is often called a \emph{behaviour}.

A behaviour is said to be nonlocal if it does not admit a decomposition of the form
\begin{equation}
P(ab|xy) = \sum_{\mu,\nu} q(\mu,\nu) D(a|x,\mu) D(b|y,\nu)  \quad \forall a,b,x,y
\end{equation}
where again $D(a|x,\mu)$ and $D(b|y,\nu)$ are deterministic probability distributions (for Alice and Bob respectively), and $q(\mu,\nu)$ is a probability distribution over $\mu$ and $\nu$ (the hidden variables), which satisfies $q(\mu,\nu)\geq 0$ $\forall \mu,\nu$ and $\sum_{\mu,\nu} q(\mu,\nu) = 1$. The decomposition \eqref{lhv} is called a local-hidden-variable (LHV) model, and any behaviour satisfying it is said to be local. Finally, for finite numbers of inputs $x$ and $y$ (and hence a finite number of deterministic strategies), determining if a behaviour is local or not can be solved by linear programming \cite{review}, a particularly simple convex optimisation problem that can be solved efficiently.

Once again we will abuse notation and denote behaviours as $P(ab|xy)$.

\subsection{Nonlocal Robustness}\label{Sec: NL robustness}

The first nonlocality quantifier we consider is the \emph{Nonlocal Robustness} ($\mathrm{NLR}$). It quantifies the minimal amount of arbitrary quantum behaviour $Q(ab|xy)$ that needs to be added to  $P(ab|xy)$ such that the mixture becomes local. Mathematically,
\begin{align}\label{NLR}
\mathrm{NLR}(P(ab|xy))& = \min\, r \\
\text{s.t. }
& \frac{P(ab|xy) + rQ(ab|xy)}{1+r} \nonumber \\
&= \sum_{\mu,\nu} q(\mu,\nu) D(a|x,\mu) D(b|y,\nu)  \:\: \forall a,b,x,y,\nonumber \\
& r\geq 0,\quad Q(ab|xy) \in \mathcal{Q}, \quad q(\mu,\nu) \geq 0. \nonumber
\end{align}
In this expression, $\mathcal{Q}$ refers to the set of quantum-realisable behaviours, \ie those that are composed of probability distributions that can be written as $Q(ab|xy)=\Tr[(M_{a|x}\otimes M_{b|y}) \rho_{\rA\rB}]$ for some choice of measurements and quantum state in some arbitrary dimension.

\subsection{Marginal Nonlocal Robustness}

Following the same lines as above, we now introduce  the \emph{Marginal Nonlocal Robustness}. This is a special instance of the quantifier introduced in the previous subsection, but now we restrict to the noise $Q(ab|xy) = P(b|y)/n$, \ie such that Alice gives outcomes with a uniform distribution $Q(a|x) = 1/n$, and Bob gives outcomes with the same marginal distribution as $P(ab|xy)$. More precisely,
\begin{align}\label{NLRMar}
\mathrm{NLR}^\mathrm{mar}(P(ab|xy))& = \min\, r \\
\text{s.t. }& r\geq 0, \quad q(\mu,\nu) \geq 0 \nonumber \\
& \frac{P(ab|xy) + rP(b|y)/n}{1+r} \nonumber \\
&= \sum_{\mu,\nu} D(a|x,\mu) D(b|y,\nu) q(\mu,\nu) \:\: \forall a,b,x,y. \nonumber
\end{align}

\subsection{LHV Nonlocal Robustness}

Similarly to above, we can also introduce the the \emph{LHV Nonlocal Robustness}. This is again a special instance of the quantifier introduced in Sec.~\ref{Sec: NL robustness}, but now we restrict the noise $Q(ab|xy)$ such that it is local. More precisely,
\begin{align}\label{NLRlhs}
\mathrm{NLR}^\mathrm{lhv}(P(ab|xy))& = \min\, r \\
\text{s.t. }& \frac{P(ab|xy) + rR(ab|xy)}{1+r} \nonumber \\
&= \sum_{\mu,\nu} q(\mu,\nu)D(a|x,\mu) D(b|y,\nu)  \:\: \forall a,b,x,y\nonumber\\
&R(ab|xy)=\nonumber\\
& \sum_{\mu,\nu} p(\mu,\nu) D(a|x,\mu) D(b|y,\nu)  \:\: \forall a,b,x,y\nonumber\\
& r\geq 0, \quad q(\mu,\nu) \geq 0,\quad p(\mu,\nu) \geq 0.\nonumber
\end{align}

\subsection{Nonlocal Weight}

We finally consider the \emph{Nonlocal Weight} \cite{EPR2} (which is also known as the EPR2 decomposition, or Nonlocal Part \cite{EPR2_Scarani, EPR2_Branciard, EPR2_Portman}). Here, a nonlocal behaviour $P(ab|xy)$ is decomposed as a convex combination of an arbitrary quantum behaviour $Q(ab|xy)$ and a local behaviour $R(ab|xy)$, with the weight of the arbitrary behaviour minimised. More precisely,
\begin{align}\label{NLW}
\mathrm{NLW}(P(ab|xy))& = \min\, r \\
\text{s.t. }& P(ab|xy) = rQ(ab|xy)+(1-r)R(ab|xy) \nonumber \\
&R(ab|xy)= \sum_{\mu,\nu} q(\mu,\nu)D(a|x,\mu) D(b|y,\nu)  \nonumber \\
&r\geq 0,\quad Q(ab|xy) \in \mathcal{Q}, \quad q(\mu,\nu) \geq 0 \nonumber
\end{align}

\section{Estimating steering and nonlocality from measurement incompatibility}\label{sec: estimating}

Having now introduced formally the relevant quantifiers of measurement incompatibility, steering, and nonlocality, we can now demonstrate our first result, which was informally given in Eq.~\eqref{e:relation}. In particular, we will show that the above-defined quantifiers satisfy the following relations:
\begin{equation}\label{e:relations}
\begin{split}
\mathrm{IR}(M_{a|x}) &\geq \mathrm{SR}(\sigma_{a|x}) \geq \mathrm{NLR}(P(ab|xy))\\
\mathrm{IR}^\mathrm{r}(M_{a|x}) &\geq \mathrm{SR}^\mathrm{red}(\sigma_{a|x}) \geq \mathrm{NLR}^\mathrm{mar}(P(ab|xy))\\
\mathrm{IR}^\mathrm{jm}(M_{a|x}) &\geq \mathrm{SR}^\mathrm{lhs}(\sigma_{a|x}) \geq \mathrm{NLR}^\mathrm{lhv}(P(ab|xy))\\
\mathrm{IW}(M_{a|x}) &\geq \mathrm{SW}(\sigma_{a|x}) \geq \mathrm{NLW}(P(ab|xy))\\
\end{split}
\end{equation}
where $\sigma_{a|x} = \Tr_\rA[(M_{a|x}\otimes\openone)\rho_{\rA\rB}]$ for any state $\rho_{\rA\rB}$, and $P(ab|xy) = \Tr[M'_{b|y}\sigma_{a|x}]$ for any set of measurements $M'_{b|y}$. That is, each incompatibility quantifier of the measurements $M_{a|x}$ upper bounds a corresponding steering quantifier of any assemblage that can be produced by Alice performing those measurements. In turn, each steering quantifier upper bounds a corresponding nonlocality quantifier of any nonlocal behaviour that Bob can produce by performing any set of measurements.

In what follows we will provide a complete demonstration for the first relation in \eqref{e:relations}. All other relations follow almost identically, and will not explicitly be shown.

We start by showing that that Incompatibility Robustness of a set of measurements $M_{a|x}$ upper bounds the Steering Robustness of any assemblage with members $\sigma_{a|x}=\Tr_\mathrm{A}[(M_{a|x}\otimes\openone) \rho_\mathrm{AB}]$, which is generated by them. Suppose the optimal solution $t^*$ for the Incompatibility Robustness \eqref{IR}, \ie $\mathrm{IR}(M_{a|x})=t^*$, is achieved by $N_{a|x}^*$. This implies that the measurements $O_{a|x}$ composed of measurement operators
\be\label{O noisy consis}
O_{a|x}=\frac{M_{a|x}+t^*N^*_{a|x}}{1+t^*}
\ee
are jointly measurable. By applying these measurements on half the state $\rho_\mathrm{AB}$ we obtain the assemblage
\be
\gamma_{a|x}=\Tr_\mathrm{A}[(O_{a|x}\otimes\openone)\rho_\mathrm{AB}]=\frac{\sigma_{a|x}+t^*\pi_{a|x}}{1+t^*}.
\ee
where $\pi_{a|x} = \Tr_\mathrm{A}[(N^*_{a|x}\otimes\openone)\rho_\mathrm{AB}]$. Since $O_{a|x}$ is jointly measurable, $\gamma_{a|x}$ is a LHS assemblage. Thus the constraints of the optimisation problem \eqref{SR} are satisfied. The pair $(t^*,\pi_{a|x})$ may not be the optimal solution of the problem \eqref{SR}, so we have that
\be\label{IRgSR}
\mathrm{IR}(M_{a|x}) = t^*\geq \mathrm{SR}(\sigma_{a|x}).
\ee
Thus, we see that, as desired, the Incompatibility Robustness provides an upper bound on the Steering Robustness of any assemblage that is created.

We now show that the Steering Robustness of an assemblage $\sigma_{a|x}$ in turn upper bounds the Nonlocal Robustness of any behaviour $P(ab|xy) = \Tr[M'_{b|y}\sigma_{a|x}] \equiv \Tr[(M_{a|x}\otimes M'_{b|y})\rho_\mathrm{AB}]$.

Similarly to before, suppose the optimal solution $s^*$ for the Steering Robustness \eqref{SR} is achieved with noise $\pi^*_{a|x}$, and consider the corresponding LHS assemblage $\gamma_{a|x}$ given by
\begin{equation}
\gamma_{a|x}=\frac{\sigma_{a|x}+s^*\pi^*_{a|x}}{1+s^*}.
\end{equation}
The behaviour that this produces, when Bob performs the measurements $M'_{b|y}$ is
\begin{align}
R(ab|xy) &= \Tr[\gamma_{a|x}M'_{b|y}] \nonumber \\
&= \frac{P(ab|xy) + s^*Q(ab|xy)}{1+s^*}
\end{align}
where $Q(ab|xy) = \Tr[\pi^*_{a|x}M'_{b|y}]$. Since $\gamma_{a|x}$ is a LHS assemblage, $R(ab|xy)$ is a local behaviour. Thus the constraints from \eqref{NLR} are satisfied and, although this is not necessarily the optimal solution, it nevertheless satisfies the relation
\begin{equation}
\mathrm{SR}(\sigma_{a|x}) = s^* \geq \mathrm{NLR}(P(ab|xy)).
\end{equation}
This shows, as desired, that the Incompatibility Robustness of a set of measurement upper bounds both the Steering Robustness of any assemblage, and the Nonlocal Robustness of any nonlocal behaviour that arises from them. By following the same line of reasoning, one sees that the same holds also for the three other families of quantifiers in \eqref{e:relations}.

\section{One-sided estimation of measurement incompatibility}\label{sec: 1sided}

In this and the following section we will consider the converse direction: Instead of looking at the measurement incompatibility as placing bounds on what can be observed in  the steering or nonlocality scenarios, we will ask how to place one-sided device-independent and device-independent bounds on the amount of measurement incompatibility. That is, we will ask, in the steering and nonlocality scenarios, how to estimate the measurement incompatibility of the (unknown) measurements of Alice.

Note that already the relations \eqref{e:relations} provide one-sided (and fully) device-independent bounds on the measurement incompatibility. However, as we will see, we can modify the steering quantifiers introduced in Sec.~\ref{sec: steering} to endow them with additional structure that will provide \emph{tighter} estimates of measurement incompability.

\subsection{Consistent Steering Robustness}
The first modified steering quantifier that we introduce is the \textit{Consistent Steering Robustness} ($\mathrm{SR}^c$), a modification of the Steering Robustness \eqref{SR}. It still quantifies the minimal amount of (arbitrary) noise $\pi_{a|x}$ that one has to add to an assemblage $\sigma_{a|x}$ such that their mixture is a LHS assemblage. The modification that is necessary to make here is to demand that the assemblage $\pi_{a|x}$ defines the \textit{same reduced state} as $\sigma_{a|x}$, \ie $\sum_{a}\sigma_{a|x}=\sum_a \pi_{a|x}=\rho_\mathrm{B}$. More precisely,
\begin{align}\label{CSR}
\mathrm{SR}^\mathrm{c}(\sigma_{a|x})=\min\, &s\\
\text{s.t. } &s\geq0,\quad \sigma_\lambda\geq0\:\: \forall \lambda,\quad \pi_{a|x}\geq0\:\:\forall x,a,\nonumber\\
&\frac{\sigma_{a|x}+s\pi_{a|x}}{1+s}=\sum_\lambda D(a|x,\lambda)\sigma_\lambda\:\: \forall a,x,\nonumber\\
&\sum_{a}\pi_{a|x}=\sum_{a}\sigma_{a|x}\:\:\forall x\nonumber.
\end{align}

Note that, since the Consistent Steering Robustness contains additional constraints compared to the the Steering Robustness, we have immediately that $\mathrm{SR}^\mathrm{c}(\sigma_{a|x}) \geq \mathrm{SR}(\sigma_{a|x})$, \ie the Consistent Steering Robustness is never smaller than the Steering Robustness.

We will now show that the Consistent Steering Robustness of an assemblage with members $\sigma_{a|x}=\Tr_\mathrm{A}[(M_{a|x}\otimes\openone) \rho_\mathrm{AB}]$ still lower bounds the Incompatibility Robustness of the measurements $M_{a|x}$ that generates this assemblage. Simiarly to before, suppose that the optimal solution for the Incompatibility Robustness \eqref{IR} is $\mathrm{IR}(M_{a|x})=t^*$, and is achieved by $N_{a|x}^*$. This implies that the measurements $O_{a|x}$ composed by measurement operators
\be\label{O noisy}
O_{a|x}=\frac{M_{a|x}+t^*N^*_{a|x}}{1+t^*}
\ee
are jointly measurable. By applying these measurements on half of the state $\rho_\mathrm{AB}$ we obtain the assemblage
\be
\gamma_{a|x}=\Tr_\mathrm{A}[(O_{a|x}\otimes\openone)\rho_\mathrm{AB}]=\frac{\sigma_{a|x}+t^*\pi_{a|x}}{1+t^*}.
\ee
where $\pi_{a|x} = \Tr_\mathrm{A}[(N^*_{a|x}\otimes\openone)\rho_\mathrm{AB}]$. Since $O_{a|x}$ is jointly measurable, $\gamma_{a|x}$ is a LHS assemblage. Moreover, $\sum_a \pi_{a|x}=\sum_a \sigma_{a|x}$ $\forall x$, since these assemblages are produced from the \textit{same} bipartite state $\rho_\mathrm{AB}$. Thus, all the constraints of the optimisation problem \eqref{CSR} are satisfied. The pair $(t^*,\pi_{a|x})$ may not be the optimal solution of the problem \eqref{CSR}, so we have that
\be\label{IRgCSR}
\mathrm{IR}(M_{a|x}) = t^*\geq \mathrm{SR}^c(\sigma_{a|x}) \geq \mathrm{SR}(\sigma_{a|x}).
\ee
Thus, with the additional constraint, we see that the Consistent Steering Robustness provides a tighter lower bound on the Incompatibility Robustness of the measurements $M_{a|x}$, performed by Alice to create the assemblage $\sigma_{a|x}$. This in particular provides a one-sided device-independent quantification of the Incompatibility Robustness.
\subsection{Consistent-LHS Steering Robustness}
Similarly to above, the \emph{Consistent-LHS Steering Robustness}, can be defined as a modification of the LHS Steering Robustness \eqref{LHSR}, where the noise added (which is a LHS assemblage) is additionally constrained to define the same reduced state as $\sigma_{a|x}$. More precisely,
\begin{align}\label{LHSRred}
\mathrm{SR}^\mathrm{c/lhs}(\sigma_{a|x})=\min\, &s\\
\text{s.t. } &s\geq0,\quad \sigma_\lambda\geq0\:\: \forall \lambda\nonumber\\
&\frac{\sigma_{a|x}+s\gamma_{a|x}}{1+s}=\sum_\lambda D(a|x,\lambda)\sigma_\lambda\:\: \forall a,x,\nonumber \\
&\gamma_{a|x}=\sum_\lambda D(a|x,\lambda)\gamma_\lambda\:\: \forall a,x,\nonumber \\
&\sum_{a}\sigma_{a|x} = \sum_{a}\gamma_{a|x} \:\:\forall x, \quad \gamma_\lambda\geq0\:\: \forall \lambda.
\end{align}
Again, since additional constraints have been added relative to \eqref{LHSR}, it follows that $\mathrm{SR}^\mathrm{c/lhs}(\sigma_{a|x}) \geq \mathrm{SR}^\mathrm{lhs}(\sigma_{a|x})$.

Following exactly the same lines as with the Consistent Steering Robustness, we can show that the Consistent-LHS Steering Robustness is again a lower bound to the Incompatibility Jointly-Measurable Robustness \eqref{jmIR} of the measurement $M_{a|x}$ that generates the assemblage $\sigma_{a|x}$, \ie
\be\label{IRlhsSRjm}
\mathrm{IR}^{jm}(M_{a|x})\geq \mathrm{SR}^\mathrm{c/lhs}(\sigma_{a|x}) \geq \mathrm{SR}^\mathrm{lhs}(\sigma_{a|x}).
\ee
%%%%%%%%%%

\subsection{Consistent Steering Weight}

The last modified quantifier we introduce is the \textit{Consistent Steering Weight}.  The difference between the standard Steering Weight \eqref{SW} and the Consistent Steering Weight is that for the latter we demand that the assemblages $\pi_{a|x}$ and $\gamma_{a|x}$ are consistent with the same reduced state as the one coming from the assemblage $\sigma_{a|x}$. Formally it is defined by:
\begin{align}
\mathrm{SW}^c(\sigma_{a|x})=\min\, &s\\
\text{s.t. } &s\geq0,\quad \pi_{a|x}\geq0\:\:\forall a,x\quad \sigma_\lambda\geq0 \:\:\forall \lambda\nonumber\\
&\sigma_{a|x}=s\pi_{a|x}+(1-s)\gamma_{a|x},\nonumber\\
& \gamma_{a|x}=\sum_\lambda D(a|x,\lambda)\sigma_\lambda\:\:\forall a,x\nonumber\\
&\sum_{a} \pi_{a|x}=\sum_{a} \gamma_{a|x} = \sum_{a} \sigma_{a|x}\:\:\forall x\nonumber
\end{align}

By following the same lines as in the previous two cases one can prove that
\be\label{IW3SWc}
\mathrm{IW}(M_{a|x})\geq \mathrm{SW}^c(\sigma_{a|x}) \geq \mathrm{SW}(\sigma_{a|x}),
\ee
for an any assemblage $\sigma_{a|x}$ that is produced by Alice performing measurements $M_{a|x}$.

Thus, we have shown how to obtain tighter one-sided device-independent bounds on measurement incompatibility be modifying previously introduced steering quantifiers in order to make them `consistent' with the additional structure that comes from the underlying measurements.

\section{Device-independent estimation of measurement incompatibility}\label{sec: DI}
In a similar fashion to the previous section, we now show that by appropriately modifying the quantifiers of nonlocality introduced in Sec.~\ref{sec: nonlocality}, we obtain device-independent lower bounds on the measurement incompatibility of either of the parties. We will present everything for the case where we estimate the incompatibility of Alice's measurements. Analogous calculations can be carried out in the case of Bob's measurements.

\subsection{Consistent Nonlocal Robustness}\label{Sec: cons NL robustness}
The \emph{Consistent Nonlocal Robustness} ($\mathrm{NLR}^c$), a modification of the Nonlocal Robustness \eqref{NLR}, quantifies the minimal amount of arbitrary quantum noise $Q(ab|xy)$ that needs to be added to a behaviour $P(ab|xy)$ such that the mixture becomes local. The modification is to further demand that the noise $Q(ab|xy)$ has a quantum realization, and that it defines the same marginal distribution for Bob as $P(ab|xy)$, \ie $Q(b|y) = P(b|y)$, where $Q(b|y) = \sum_a Q(ab|xy)$ which is independent of $x$ due to no-signalling. Mathematically,
\begin{align}\label{NLRc}
\mathrm{NLR}^c(P(ab|xy))& = \min\, r \\
\text{s.t. }& r\geq 0,\quad Q(ab|xy) \in \mathcal{Q}, \quad q(\mu,\nu) \geq 0 \nonumber \\
& \frac{P(ab|xy) + rQ(ab|xy)}{1+r} \nonumber \\
&= \sum_{\mu,\nu} D(a|x,\mu) D(b|y,\nu) q(\mu,\nu) \:\: \forall a,b,x,y,\nonumber \\
& Q(b|y) = P(b|y) \:\: \forall b,y,
\end{align}
where $\mathcal{Q}$ is the set of quantum-realisable behaviours. Just as in the case of steering, since the Consistent Nonlocal Robustness contains additional constraints compared to the Nonlocal Robustness, it immediately follows that $\mathrm{NLR}^\mathrm{c}(P(ab|xy)) \geq \mathrm{NLR}(P(ab|xy))$.

We now show that the Consistent Nonlocal Robustness of the behaviour defined by $P(ab|xy) = \Tr[(M_{a|x}\otimes M'_{b|y})\rho_\mathrm{AB}]$ lower bounds the Consistent Steering Robustness of the assemblage $\sigma_{a|x} = \Tr_\mathrm{A}[(M_{a|x}\otimes \openone)\rho_\mathrm{AB}]$ that Alice prepares for Bob using measurements $M_{a|x}$. Since the Consistent Steering Robustness lower bounds the Incompatibility Robustness, we thus obtain a device-independent lower bound on the Incompatibility Robustness (and on the one-sided device-independent Consistent Steering Robustness).

Similarly to previously, suppose the optimal solution $s^*$ for the Consistent Steering Robustness \eqref{CSR} is achieved with noise $\pi^*_{a|x}$, and consider the corresponding LHS assemblage $\gamma_{a|x}$ given by
\begin{equation}
\gamma_{a|x}=\frac{\sigma_{a|x}+s^*\pi^*_{a|x}}{1+s^*}.
\end{equation}
The behaviour that this produces, when Bob performs the measurements $M'_{b|y}$ is
\begin{align}
R(ab|xy) &= \Tr[\gamma_{a|x}M'_{b|y}] \nonumber \\
&= \frac{P(ab|xy) + s^*Q(ab|xy)}{1+s^*}
\end{align}
where $Q(ab|xy) = \Tr[\pi^*_{a|x}M'_{b|y}]$. Since $\gamma_{a|x}$ is LHS, $R(ab|xy)$ is a local behaviour. Furthermore, we have $\sum_a P(ab|xy) = P(b|y) = Q(b|y) = \sum_a Q(ab|xy)$, since due to the consistency requirement $\sum_a \sigma_{a|x} = \sum_a \pi^*_{a|x}$. Thus all the constraints from \eqref{NLRc} are satisfied and, although this is not necessarily the optimal solution, it nevertheless provides the lower bound
\begin{equation}
\mathrm{IR}(M_{a|x})
 \geq \mathrm{SR}^c(\sigma_{a|x})\geq \mathrm{NLR}^c(P(ab|xy)).
\end{equation}

%\subsection{Marginal Nonlocal Robustness}
%Following the same lines as above, we now introduce the next modified nonlocal quantifier, the \emph{Marginal Nonlocal Robustness}. This is again a special instance of the quantifier introduced in the previous subsection, but now we restrict to the noise $Q(ab|xy) = P(b|y)/n$, \ie such that Alice gives outcomes with a uniform distribution $Q(a|x) = 1/n$, and Bob gives outcomes with the same marginal distribution as $P(ab|xy)$. More precisely,
%\begin{align}\label{NLRMar}
%\mathrm{NLR}^\mathrm{mar}(P(ab|xy))& = \min\, r \\
%\text{s.t. }& r\geq 0, \quad q(\mu,\nu) \geq 0 \nonumber \\
%& \frac{P(ab|xy) + rP(b|y)/n}{1+r} \nonumber \\
%&= \sum_{\mu,\nu} D(a|x,\mu) D(b|y,\nu) q(\mu,\nu) \:\: \forall abxy. \nonumber
%\end{align}
%Once again, following exactly the same line of reasoning as in the previous examples, it is straightforward to show that the Marginal Nonlocal Robustness of the behaviour $P(ab|xy) = \Tr[(M_{a|x}\otimes M'_{b|y})\rho_\mathrm{AB}]$ provides a lower bound to the Reduced-state Steering Robustness of the assemblage $\sigma_{a|x} = \Tr_\mathrm{A}[(M_{a|x}\otimes \openone)\rho_\mathrm{AB}]$ which in turn provides a lower bound on the Incompatibility Random Robustness of the set of measurements $M_{a|x}$, that is
%\begin{equation}
%\mathrm{IR}^r(M_{a|x})
% \geq \mathrm{SR}^\mathrm{red}(\sigma_{a|x})\geq \mathrm{NLR}^\mathrm{mar}(P(ab|xy)).
% \end{equation}

 \subsection{Consistent-LHV Nonlocal Robustness}

Similarly to before, we also introduce the the \emph{consistent-LHV Nonlocal Robustness}. This is a modification of \eqref{LHSR}, (where the noise $Q(ab|xy)$ must be local, and like above, we now demand that it has the same marginal distribution as $P(ab|xy)$). More precisely,
\begin{align}\label{NLRlhsc}
\mathrm{NLR}^\mathrm{c/lhs}(P(ab|xy))& = \min\, r \\
\text{s.t. }& r\geq 0, \quad q(\mu,\nu) \geq 0 \nonumber \\
& \frac{P(ab|xy) + rQ(ab|xy)}{1+r} \nonumber \\
&= \sum_{\mu,\nu} q(\mu,\nu) D(a|x,\mu) D(b|y,\nu)  \:\: \forall abxy\nonumber\\
&Q(ab|xy)=\nonumber\\
& \sum_{\mu,\nu} p(\mu,\nu) D(a|x,\mu) D(b|y,\nu)  \:\: \forall abxy\nonumber\\
&p(\mu,\nu) \geq 0,\quad
Q(b|y) = P(b|y) \:\: \forall b,y. \nonumber
\end{align}
Once again, it is straightforward to show that the consistent-LHV Nonlocal Robustness of the behaviour $P(ab|xy) = \Tr[(M_{a|x}\otimes M'_{b|y})\rho_\mathrm{AB}]$ lower bounds the Consistent LHS Steering Robustness of the assemblage $\sigma_{a|x} = \Tr_\mathrm{A}[(M_{a|x}\otimes \openone)\rho_\mathrm{AB}]$ which was shown before to lower bound on the Incompatibility Jointly-Measurable Robustness of the set of measurements $M_{a|x}$:
\begin{equation}
\mathrm{IR}^{jm}(M_{a|x})
 \geq \mathrm{SR}^\mathrm{c/lhs}(\sigma_{a|x})\geq \mathrm{NLR}^\mathrm{c/lhv}(P(ab|xy)).
 \end{equation}

\subsection{Consistent Nonlocal Weight}
We finally consider the quantifier the \emph{Consistent Nonlocal Weight}, a modification of the Nonlocal Weight \eqref{NLW}. We add the additional constraint that both parts of the decomposition, $Q(ab|xy)$ and $R(ab|xy)$, must have the same marginal distribution as the original behaviour $P(ab|xy)$. More precisely,
\begin{align}\label{NLWc}
\mathrm{NLW}^c(P(ab|xy))& = \min\, r \\
\text{s.t. }& r\geq 0,\quad Q(ab|xy) \in \mathcal{Q}, \quad q(\mu,\nu) \geq 0 \nonumber \\
& P(ab|xy) = rQ(ab|xy)+(1-r)R(ab|xy) \nonumber \\
&R(ab|xy)= \sum_{\mu,\nu} D(a|x,\mu) D(b|y,\nu) q(\mu,\nu) \nonumber \\
& Q(b|y) = R(b|y) = P(b|y) \:\: \forall b,y.
\end{align}
As in all previous cases, it is straightforward to show that the Consistent Nonlocal Weight of the behaviour $P(ab|xy) = \Tr[(M_{a|x}\otimes M'_{b|y})\rho_\mathrm{AB}]$ provides a lower bound to the Consistent Steering Weight of the assemblage $\sigma_{a|x} = \Tr_\mathrm{A}[(M_{a|x}\otimes \openone)\rho_\mathrm{AB}]$ which in turn provides a lower bound on the Incompatibility Weight of the set of measurements $M_{a|x}$, that is
\begin{equation}
\mathrm{IW}(M_{a|x})\geq \mathrm{SW}^c(\sigma_{a|x})\geq\mathrm{NLW}^c(P(ab|xy))
\end{equation}

Thus, in all cases we can obtain device-independent bounds on the `consistent' quantifiers of steering, which themselves were designed specifically to give tighter one-sided device-independent bounds on useful quantifiers of measurement incompatibility.

\section{Tightness for entangled pure states}\label{sec: pure states}

We now show that the one-sided device-independent lower bounds provided by the `consistent' quantifiers are in fact tight if the state shared between Alice and Bob is a full Schmidt-rank pure entangled state. This justifies the modifications presented, and shows that they cannot be further improved. The main ingredient we are going to use is that by measuring half of a pure entangled state $\ket{\psi}$ we obtain the following assemblage:
\ba\label{pure-state ass}
\sigma_{a|x}&=&\Tr_\mathrm{A}[(M_{a|x}\otimes\openone) \ketbra{\psi}{\psi}]\nonumber\\
&=&\rho_\mathrm{B}^{\frac{1}{2}}M_{a|x}^\mathrm{T}\rho_\mathrm{B}^{\frac{1}{2}},
\ea
where again $\rho_\mathrm{B}=\sum_a\sigma_{a|x}$ is the reduced state of $\ketbra{\psi}{\psi}$, $\rho_\mathrm{B} = \Tr_\mathrm{A}[\ketbra{\psi}{\psi}]$ and $\mathrm{T}$ denotes transposition (in the eigenbasis of $\rho_\mathrm{B}$).

In what follows we will show that the Consistent Steering Robustness of the assemblage \eqref{pure-state ass} equals the Incompatibility Robustness of $M_{a|x}$. Consider an optimal solution $(s^*, \pi_{a|x}^*,\sigma_\lambda^*)$ for the problem  \eqref{CSR}, \ie $\mathrm{SR}^c(\sigma_{a|x})=s^*$, such that
\ba
\gamma_{a|x}&=&\frac{\sigma_{a|x}+s^*\pi_{a|x}^*}{1+s^*},\nonumber\\
&=&\frac{\rho_\mathrm{B}^{\frac{1}{2}}M_{a|x}^\mathrm{T}\rho_\mathrm{B}^{\frac{1}{2}}+s^*\rho_\mathrm{B}^{\frac{1}{2}}N_{a|x}^\mathrm{T}\rho_\mathrm{B}^{\frac{1}{2}}}{1+s^*},\nonumber\\
&=&\sum_\lambda D(a|x,\lambda)\sigma^*_\lambda,
\ea
where in the second equality we used \eqref{pure-state ass}, the fact that $\pi^*_{a|x}$ can be obtained by some measurements $N_{a|x}$ on the same state $\ketbra{\psi}{\psi}$, and in the third equality that $\gamma_{a|x}$ has a LHS model. By multiplying this equation by $\rho_\mathrm{B}^{-\frac{1}{2}}$ from both sides and applying the transposition map we obtain
\ba\label{JM structure}
\frac{M_{a|x}+s^*N_{a|x}^*}{1+s^*}&=&\left(\sum_\lambda D(a|x,\lambda)\rho_\mathrm{B}^{-\frac{1}{2}}\sigma^*_\lambda\rho_\mathrm{B}^{-\frac{1}{2}}\right)^\mathrm{T}\nonumber\\
&=&\sum_\lambda D(a|x,\lambda)G_\lambda,
\ea
where $G_\lambda=(\rho_\mathrm{B}^{-\frac{1}{2}}\sigma^*_\lambda\rho_{R}^{-\frac{1}{2}})^\mathrm{T}$.
Notice that the right-hand-side defines a valid set of measurements, since they are positive operators that sum up to the identity:
\ba
&&\sum_a\sum_\lambda D(a|x,\lambda)G_\lambda\nonumber\\
&=&\sum_a \left(\sum_\lambda D(a|x,\lambda)\rho_\mathrm{B}^{-\frac{1}{2}}\sigma^*_\lambda\rho_\mathrm{B}^{-\frac{1}{2}}\right)^\mathrm{T}\nonumber\\
&=&\left(\sum_\lambda \rho_\mathrm{B}^{-\frac{1}{2}}\sigma^*_\lambda\rho_\mathrm{B}^{-\frac{1}{2}}\right)^\mathrm{T}=\left(\rho_\mathrm{B}^{-\frac{1}{2}}\sum_\lambda  \sigma^*_\lambda\rho_\mathrm{B}^{-\frac{1}{2}}\right)^\mathrm{T}\nonumber\\
&=&(\rho_\mathrm{B}^{-\frac{1}{2}}\rho_\mathrm{B}\rho_\mathrm{B}^{-\frac{1}{2}})^\mathrm{T}=\openone.
\ea
Furthermore Eq. \eqref{JM structure} says that
$$\frac{M_{a|x}+s^*N_{a|x}^*}{1+s^*}$$ is jointly measurable, which implies that
\be
\mathrm{IR}(M_{a|x})\leq s^*=\mathrm{SR}^c(\sigma_{a|x}),
\ee
which, together with \eqref{IRgCSR} implies that $\mathrm{IR}(M_{a|x})=\mathrm{SR}^c(\sigma_{a|x})$.

Similarly one has that $\mathrm{IR}^r(M_{a|x})=\mathrm{SR}^\mathrm{red}(\sigma_{a|x})$ whenever $\sigma_{a|x}=\Tr_\mathrm{A}[(M_{a|x}\otimes\openone) \ketbra{\psi}{\psi}]$ and $\ket{\psi}$ is a full Schmidt-rank state. Again consider an optimal solution $(s^*,\sigma_\lambda^*)$ such that $\mathrm{SR}^r(\sigma_{a|x})=s^*$,
\ba
\frac{\sigma_{a|x}+s^*\rho_\mathrm{B}/n}{1+s^*}
&=&\frac{\rho_\mathrm{B}^{\frac{1}{2}}M_{a|x}^\mathrm{T}\rho_\mathrm{B}^{\frac{1}{2}}+s^*\rho_\mathrm{B}/n}{1+s^*}\nonumber\\
&=&\sum_\lambda D(a|x,\lambda)\sigma^*_\lambda.
\ea
Multiplying by $\rho_\mathrm{B}^{-\frac{1}{2}}$ from both sides and applying the transposition map we obtain
\ba
\frac{M_{a|x}+s^*\openone/n}{1+s^*}&=&\left(\sum_\lambda D(a|x,\lambda)\rho_\mathrm{B}^{-\frac{1}{2}}\sigma^*_\lambda\rho_\mathrm{B}^{-\frac{1}{2}}\right)^\mathrm{T}\nonumber\\
&=&\sum_\lambda D(a|x,\lambda)G_\lambda,
\ea
where again $G_\lambda=(\rho_\mathrm{B}^{-\frac{1}{2}}\sigma^*_\lambda\rho_{R}^{-\frac{1}{2}})^\mathrm{T}$. This implies that $\mathrm{IR}^r(M_{a|x})\leq \mathrm{SR}^\mathrm{red}(\sigma_{a|x})$, which together with \eqref{e:relations} gives $\mathrm{IR}^r(M_{a|x})=\mathrm{SR}^\mathrm{red}(\sigma_{a|x})$.

By following the same line of reasoning one can also show that $\mathrm{IR}^\mathrm{jm}(M_{a|x})=\mathrm{SR}^\mathrm{c/lhs}(\sigma_{a|x})$ whenever the measured state is a full Schmidt-rank pure state.

We can finally show that $\mathrm{IW}(M_{a|x})=\mathrm{SW}^c(\sigma_{a|x})$ whenever $\sigma_{a|x}=\Tr_\mathrm{A}[(M_{a|x}\otimes\openone) \ketbra{\psi}{\psi}]$. Again consider an optimal solution $(s^*,\pi_{a|x}^*,\sigma_\lambda^*)$ such that $\mathrm{SW}^c(\sigma_{a|x})=s^*$, and therefore
\ba
\sigma_{a|x}=s^*\pi^*_{a|x}+(1-s^*)\sum_\lambda D(a|x,\lambda)\sigma^*_\lambda.
\ea
Using \eqref{pure-state ass} we have that
\be
\rho_\mathrm{B}^{\frac{1}{2}}M_{a|x}^\mathrm{T}\rho_\mathrm{B}^{\frac{1}{2}}=s^*\rho_\mathrm{B}^{\frac{1}{2}}O_{a|x}^\mathrm{T}\rho_\mathrm{B}^{\frac{1}{2}}+(1-s^*)\sum_\lambda D(a|x,\lambda)\sigma^*_\lambda.
\ee
Multiplying this equation by $\rho_\mathrm{B}^{-\frac{1}{2}}$ from both sides and applying the transposition map we obtain
\ba
M_{a|x}&=&s^*O_{a|x}+(1-s^*)\left(\sum_\lambda D(a|x,\lambda)\rho_\mathrm{B}^{-\frac{1}{2}}\sigma^*_\lambda\rho_\mathrm{B}^{-\frac{1}{2}}\right)^\mathrm{T}\nonumber\\
&=&s^*O_{a|x}+(1-s^*)\sum_\lambda D(a|x,\lambda)G_\lambda.
\ea
This implies that $\mathrm{IW}(M_{a|x})\leq \mathrm{SW}^{c}(\sigma_{a|x})$, which together with \eqref{IW3SWc} gives $\mathrm{IW}(M_{x})=\mathrm{SW}^{c}(\sigma_{a|x})$.

In summary, in all three cases, whenever the parties share a full Schmidt-rank pure state, we see that the consistent steering quantifier exactly quantifies the corresponding amount of incompatibility.

It is not straightforward to obtain tightness with respect to the nonlocality measures, since in this case the measurements performed by Bob also plays a role in the bound. An interesting question is whether for every set of measurements chosen for Alice, whether one can always find a bipartite quantum state, and a set of measurements for Bob, in order to saturate the previous bounds.

\section{Examples}\label{sec: examples}
In order to demonstrate the theory presented in this paper, we will first analyse theoretical examples for the two-qubit Werner state and pure partially-entangled states before moving on to analyse the recent loophole-free demonstrations of steering \cite{LFsteer1,LFsteer2} and nonlocality \cite{LFnonl1,LFnonl2,LFnonl3}, to calculate the bounds that these experiments placed upon the incompatibility of the measurements used. Our numerical calculations were performed using the package {\sc cvx} for {\sc matlab} \cite{cvx} with the solver {\sc mosek} \cite{mosek}. In addition, the packages {\sc qetlab} \cite{qetlab} and {\sc the tensor toolbox} \cite{tensor} were utilised.
Notice that the set $\mathcal{Q}$ has no known simple characterisation. However, by using the outer approximations to $\mathcal{Q}$ provided by the Navascu\'es-Pironio-Acin (NPA) hierarchy of SDP relaxations \cite{NPA1,NPA2}, one can obtain lower bounds on $\mathrm{NLR}^c$ and $\mathrm{NLR}$ by means of solving a SDP. In the examples below, the second level, $\mathcal{Q}^{2}$, of the NPA hierarchy was used.

\subsection{Theoretical two-qubit examples}
\begin{figure}[h]
\begin{center}
\includegraphics[width=\columnwidth]{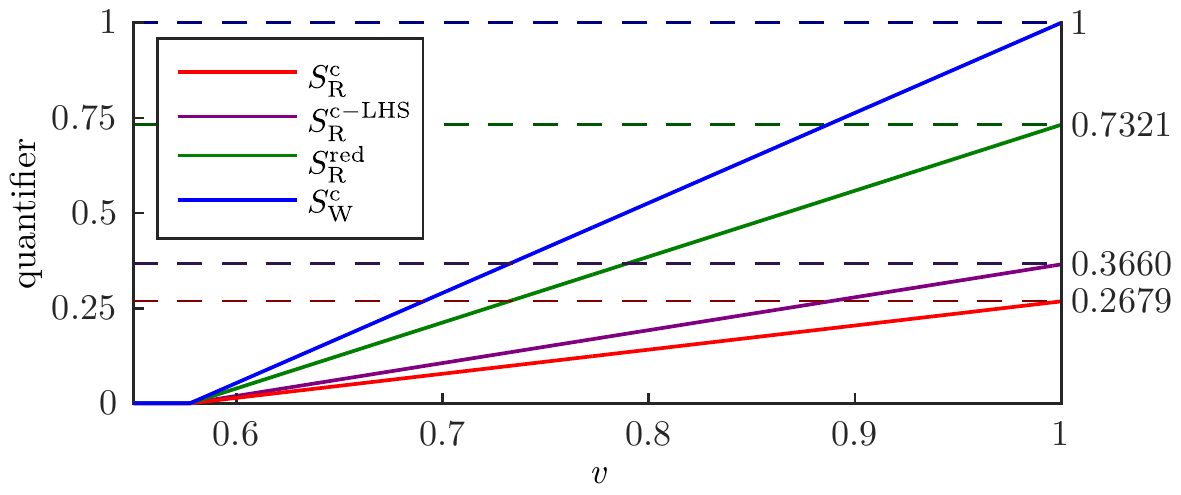}
\caption{\label{SvsV} Plot of consistent steering quantifiers vs.~visibility $V$ of two-qubit Werner states \eqref{LFSstate1}. Alice performs measurements of Pauli operators $X$, $Y$ and $Z$. As can be seen, each quantifier becomes non-zero when $v = 1/\sqrt{3}$, the critical visibility for demonstrating steering with three measurements,  and grows linearly with $v$. The dashed lines, and values on the right-hand axis, show the values for the corresponding incompatibility quantifiers.
}

\includegraphics[width=\columnwidth]{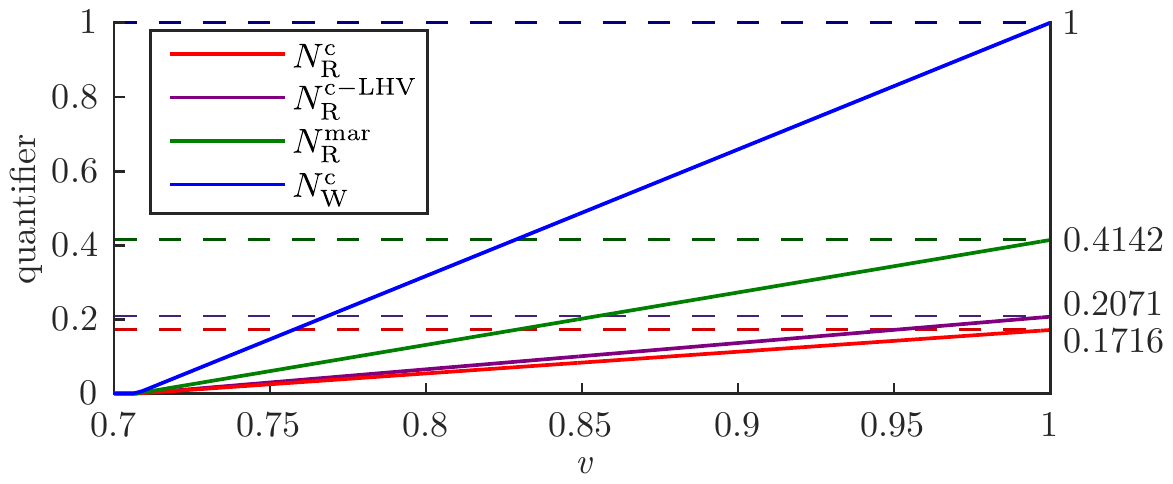}
\caption{\label{NLvV} Plot of consistent nonlocality quantifiers vs.~visibility $v$ of two-qubit Werner states \eqref{LFSstate1}. Alice performs measurements of Pauli operators $X$ and $Z$, while Bob performs measurements of $(X+Z)/\sqrt{2}$ and $(X-Z)/\sqrt{2}$, \ie the optimal measurements for violating the CHSH inequality. As can be seen, each quantifier becomes non-zero when $v = 1/\sqrt{2}$, the critical visibility for demonstrating nonlocality with two measurements, and grows linearly with $v$. The dashed lines, and values on the right-hand axis, show the values for the corresponding incompatibility quantifiers.
}
\end{center}
\end{figure}

We start by considering that the state shared between Alice and Bob is the two-qubit Werner state
\begin{equation}\label{LFSstate1}
\rho_\mathrm{AB} = v\ket{\psi}\bra{\psi} + (1-v)\openone/4,
\end{equation}
where $\ket{\psi} = (\ket{00}+\ket{11})/\sqrt{2}$ is a maximally entangled state, and $v$ is the visibility. We consider the bounds that can be placed on the measurement incompatibility of Alice's measurements in both the steering and nonlocality scenarios. In the steering scenario, Alice performs the three mutually unbiased Pauli measurements $X$, $Y$, and $Z$. In the nonlocality scenario, Alice performs $X$ and $Z$ measurements, while Bob performs measurements of $(X+Z)/\sqrt{2}$ and $(X-Z)/\sqrt{2}$, \ie the optimal measurements for violating the CHSH inequality. The results are summarised in Fig.~\ref{SvsV} and Fig.~\ref{NLvV} respectively. As can be seen, in all cases the lower bounds happen to be linear functions of the visibility $v$.

\begin{figure}[h]
\begin{center}
\includegraphics[width=\columnwidth]{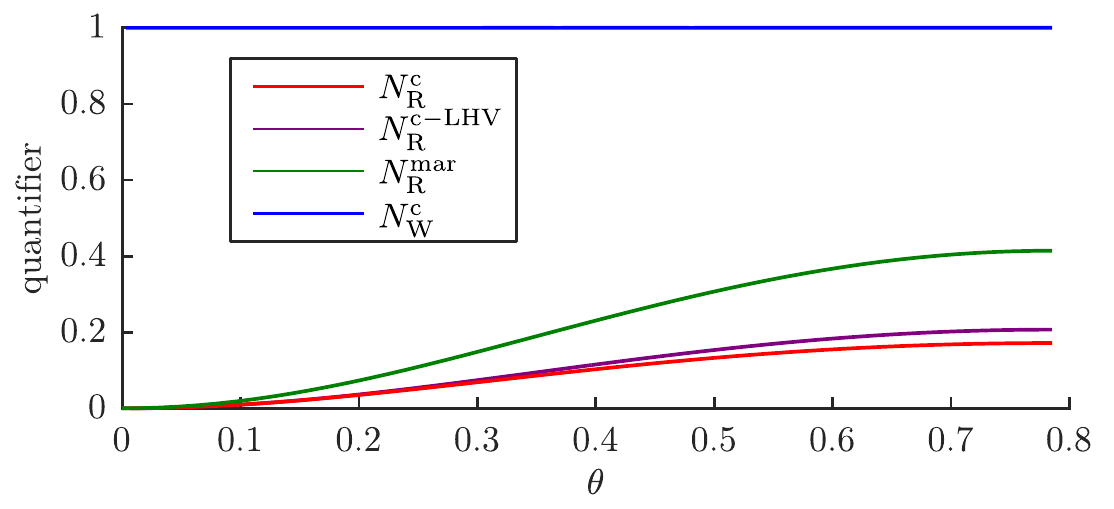}
\caption{\label{NLvstheta} Plot of consistent nonlocality quantifiers vs. $\theta$ of two-qubit partially-entangled state. Alice performs measurements of Pauli operators $X$ and $Z$, while Bob performs two measurements that are numerically optimised for each quantifier and each value of $\theta$. As can be seen, each quantifier increases monotonically with $\theta$. Moreover, for the Consistent Nonlocal Weight, for the entire parameter range it takes the maximum value $\mathrm{NLW}^c = 1$. The limiting values at $\theta = \pi/4$ coincide with those from Fig.~\ref{NLvV}, and are equal to the corresponding values of the incompatibility quantifiers for the measurements.
}
\end{center}
\end{figure}

We next consider the device-independent lower bounds in the case that a pure partially-entangled state is distributed between Alice and Bob, $\ket{\phi} = \cos \theta \ket{00} + \sin \theta \ket{11}$, for $\theta \in (0,\pi/4]$. We only consider nonlocality, since in the previous section we showed that consistent steering quantifiers are equal to incompatibility quantifiers for the case of pure (full Schmidt-rank) entangled states. We keep Alice's measurements fixed at $X$ and $Z$, but optimise the measurements performed by Bob, with the only restriction that Bob performs two measurements. For the Consistent Nonlocal Robustness, and the Marginal Nonlocal Robustness we perform a heuristic see-saw optimisation to search for optimal POVM measurements for each value of $\theta$. For the Consistent Nonlocal Weight, we use the results of \cite{AciMasPir12} to conclude that it is unity for all values of $\theta \in (0,\pi/4]$\footnote{In particular, \cite{AciMasPir12} present a Bell inequality $I_\alpha$, which is \textit{maximally} violated by partially entangled states when Alice measures $X$ and $Z$, and Bob measures $\cos\mu Z + \sin \mu X$ and $\cos \mu Z - \sin \mu X$, where $\tan \mu = \sin 2\theta/\alpha$. The fact that the violation is maximal implies that the Consistent Nonlocal Weight is unity. Indeed, denoting the maximal quantum violation $\beta_\mathcal{Q}$, and the local bound $\beta_\mathrm{LHV}$, then the $I_\alpha(P(ab|xy)) = r I_\alpha(Q(ab|xy)) + (1-r)I_\alpha(R(ab|xy)) \leq r\beta_\mathcal{Q} + (1-r)\beta_\mathrm{LHV}$. Re-arranging gives $r \geq (I_\alpha(P(ab|xy)) - \beta_\mathrm{LHV})/(\beta_\mathcal{Q} - \beta_\mathrm{LHV})$. Thus, if $I_\alpha(P(ab|xy)) = \beta_\mathcal{Q}$, then $r \geq 1$. Finally, the consistency constraints become trivial (since $P(ab|xy) = Q(ab|xy)$ and $(1-r)R(ab|xy) = 0$).}. The results are displayed in Fig.~\ref{NLvstheta}.

\subsection{Loophole-free steering demonstrations}
We start by analysing the recent loophole-free demonstrations of steering that were reported in \cite{LFsteer1,LFsteer2}. In the experiment of Wittmann \etal\ \cite{LFsteer1}, polarization-entangled photons were distributed between Alice and Bob in the state $\rho_\mathrm{AB}$, with both parties performing measurements of the three Pauli operators $X$, $Y$ and $Z$ (labelled by $x = 0,1,2$ respectively). Due to losses, both on the channel and at the detectors, instead of performing ideal dichotomic measurements (where the measurement operators are projectors $\Pi_{a|x}$, for $a = 0,1$), Alice performs lossy POVM measurements with an additional third outcome $\nc$ (corresponding to `no-click' events), with POVM elements
\begin{align}\label{LFSmeas1}
M_{a|x} =
  \begin{cases}
   \eta_x \Pi_{a|x} &\text{for } a=0,1 \\
   (1-\eta_x)\openone &\text{for } a=\nc
  \end{cases}
\end{align}
where $\eta_x$ quantifies the amount of loss for the measurement labelled by $x$. In \cite{LFsteer1}, Wittmann \etal  report $(\eta_0,\eta_1,\eta_2) = (0.382, 0.383, 0.383)$. Furthermore, it is stated that the state produced is of the form \eqref{LFSstate1}, where $\ket{\psi} = (\ket{HV}-\ket{VH})\sqrt{2}$ is the maximally entangled state of polarization (where $H$ and $V$ represent horizontal and vertical polarization, respectively) and $v = 0.9556$.

We used this information to evaluate (i) the incompatibility quantifiers corresponding to the theoretical description of the measurements performed (using as input the experimentally observed efficiencies) and (ii) the one-sided device-independent bounds, using the theoretical assemblage that would arise by applying the measurements \eqref{LFSmeas1} onto the state \eqref{LFSstate1}, utilising the experimentally observed inefficiencies and visibility. The available data does not allow us to infer the experimental assemblages, which is why we calculated theoretically the assemblage based upon the experimentally observed data. The results are summarised in the first two rows of Table \ref{LFStable}.
\begin{table}[t]
    \begin{tabular}{c|c|c|c}
     & $\mathrm{IR}$ &$\mathrm{IR}^r$ & $\mathrm{IW}$ \\ \hline
    Wittmann \etal & $1.204\times 10^{-2}$  & $4.112\times 10^{-2}$ &  $4.963\times 10^{-2}$ \\
        Bennet \etal & $1.841\times 10^{-3}$  & $5.840\times 10^{-3}$ &  $3.556\times 10^{-2}$
    \end{tabular}
\medskip \,
    \begin{tabular}{c|c|c|c}
     & $\mathrm{SR}^c$ &$\mathrm{SR}^\mathrm{red}$ & $\mathrm{SW}^c$ \\ \hline
    Wittman \etal & $7.406\times 10^{-3}$  & $2.528\times 10^{-2}$ &  $3.052\times 10^{-2}$ \\
    Bennett \etal & $1.283\times 10^{-3}$  & $4.071\times 10^{-3}$ &  $2.228\times 10^{-2}$
    \end{tabular}
    \caption{\label{LFStable} Incompatibility quantifiers and associated one-sided device-independent Consistent steering quantifiers for the measurements used in the loophole-free steering demonstrations of Wittmann \etal\ \cite{LFsteer1} and Bennet \etal\  \cite{LFsteer2}.}
\end{table}

In the experiment of Bennet \etal  \cite{LFsteer2} again polarization-entangled photons were distributed between Alice and Bob in the state \eqref{LFSstate1} but now with the visibility $v = 0.992$. Here, Alice made 10 measurements (of the form \eqref{LFSmeas1}), with Bloch vectors pointing in the direction of the vertices of a dodecahedron\footnote{the 20 vertices consist of ten pairs of antipodal vertices}, and average loss $\eta = 0.132$. Similarly to above, this data allows for a theoretical analysis of the measurement incompatibility and one-sided device-independent bounds. The results are summarised in the last two rows of Table \ref{LFStable}.

It is evident that in both experiments, due to the amount of loss in the systems, the quantifiers are small, but necessarily non-zero since steering could not have been observed without incompatible measurements. Moreover, due to the high-quality nature of the sources, the one-sided device-independent bounds are relatively tight.

\subsection{Loophole-free nonlocality demonstrations}
We now move on to the very recent loophole-free demonstrations of nonlocality presented in \cite{LFnonl1,LFnonl2,LFnonl3}. In all three cases we used the data available to reproduce the experimentally observed behaviour $P_\mathrm{exp}(ab|xy)$. Thus in the following we will bypass having to make assumptions about the underlying states and measurements used. The only difficulty that arises is that the experimentally observed behaviours $P_\mathrm{exp}(ab|xy)$ will never strictly satisfy the no-signalling conditions -- since these are linear \emph{equality} constraints, they will always be violated due to the finite-size statistics used to estimate the underlying probabilities. As such, they cannot be directly used in the programs \eqref{NLRc}, \eqref{NLRMar} and \eqref{NLWc}, which assume input data that is strictly no-signalling\footnote{Note that the same problem would have arisen in the previous case if we had had access to the experimentally observed assemblages.}. To overcome this problem, we first find the non-signalling behaviour that most closely approximates the experimental behaviour, and use this as our `best estimate' of the true underlying probability distribution\footnote{More precisely, we find $P_\mathrm{NS}^* = \argmin_{P_\mathrm{NS}} D(P_\mathrm{exp}||P_\mathrm{NS})$, the non-signalling behaviour that minimises the relative-entropy with respect to the experimental behaviour. Note that this minimisation problem, although not an SDP itself, is solved natively by {\sc cvx} using it's successive approximation technique \cite{cvx}. That is, the problem is handled by solving a sequence of SDPs, which yields the solution to the full accuracy of the core solver.}

\begin{table}[t]
    \begin{tabular}{c|c|c|c}
     & $\mathrm{NLR}^c$ & $\mathrm{NLR}^\mathrm{mar}$ & $\mathrm{NLW}^c$ \\ \hline
    Hensen \etal \ & $9.097\times 10^{-2}$  & $2.194\times 10^{-1}$ &  $5.330\times 10^{-1}$ \\
        %Hensen \etal \ (B)& $9.088\times 10^{-2}$  & $2.194\times 10^{-1}$ &  $5.299\times 10^{-1}$ \\ \hline
    Giustina \etal \ & $6.681\times 10^{-6}$  & $1.339\times 10^{-5}$ &  $3.345\times 10^{-2}$ \\
        %Guistina \etal \ (B)& $6.675\times 10^{-6}$  & $1.339\times 10^{-5}$ &  $3.250\times 10^{-2}$ \\ \hline
    Shalm \etal \ & $1.192\times 10^{-5}$  & $2.389\times 10^{-5}$ &  $6.546\times 10^{-2}$ \\
        %Shalm \etal \ (B)& $1.192\times 10^{-5}$  & $2.389\times 10^{-5}$ &  $6.508\times 10^{-2}$
    \end{tabular}

    \caption{\label{LFNLtable} Consistent Nonlocality quantifiers for the measurements used by Alice in the loophole-free Bell nonlocality demonstrations of Hensen \etal\ \cite{LFnonl1}, Giustina \etal\ \cite{LFnonl2} and Shalm \etal\ \cite{LFnonl3}.}
\end{table}

In the experiment of Hensen \etal \cite{LFnonl1} a so-called `event-ready scheme' was used that, through entanglement swapping, generated entanglement between distant
electron spins which were then measured using spin read-out measurements. In the experiments of Giustina \etal\cite{LFnonl2} and Shalm \etal \cite{LFnonl3}, polarization-entangled photons were used, along with polarization measurements. In all three experiments both parties performed two measurements with two outcomes (no-click outcomes were binned along with the one of the outcomes of the ideal measurements). In Table \ref{LFNLtable} the device-independent bounds on the incompatibility of the measurements of Alice in each experiment are given.

\section{Conclusions}\label{sec: conclusion}
In this paper we have shown a quantitative relation between measurement incompatibility, steering and nonlocality quantifiers. This allows to estimate how much steering and nonlocality a set of measurements can provide, or to place one-sided device-independent and device-independent bounds on measurement incompatibility. In particular, we showed how, for a number of incompatibility quantifiers, one can modify associated steering and nonlocality quantifiers, such that the data obtained in a steering or Bell test tightly bounds the incompatibility of the measurements performed by the uncharacterised devices. As an example of our technique, we studied the data from recent loophole-free demonstrations of steering and nonlocality, so showed how much measurement incompatibility these experiments certified.

Although we only presented our analysis for four example quantifiers, it is clear from the general method of constructing `consistent' quantifiers that the technique works more generally (in particular it is clear that it works for any quantifier defined through a similar variational principle. We finally note that the dual programs of the SDPs defining the new steering and nonlocality quantifiers presented here will provide, respectively, new linear steering and Bell inequalities.

Finally, an interesting direction for future work would be to look at the case of continuous variable (CV) systems. There, one can no longer work directly with assemblages or nonlocal behaviours, due to the infinite dimensional nature of the systems involves, and the continuous measurement outcomes. Nevertheless, it would be interesting to see if the recently introduced quantifiers of CV steering \cite{CV1,CV2} provide a quantitative relation to CV measurement incompatibility.

\emph{Note added:} After uploading a preprint of this work to the arXiv, we learnt of independent work of Chen, Budroni and Liang, that also provides a lower-bound on the  Incompatibility Robustness in a device-independent manner \cite{CBL}.

\begin{acknowledgements}
We thank M. T. Quintino and N. Brunner for useful discussions. This work was supported by the Beatriu de Pinos fellowship (BP-DGR-2013), Spanish MINECO (Severo Ochoa grant SEV-2015-0522), the AXA Chair in Quantum Information Science, and the ERC AdG NLST.
\end{acknowledgements}

\end{document}